\documentclass{article}\usepackage{jkas2}
\runningauthor{R.Cassano \& G.Brunetti}
\runningtitle{Cluster Mergers and Non-Thermal Phenomena:
A Statistical Magneto-Turbulent Model}
\beginpage{1}
\endpage{5}

\begin{document}

\font\twelvei = cmmi10 scaled\magstep1 
       \font\teni = cmmi10 \font\seveni = cmmi7
\font\mbf = cmmib10 scaled\magstep1
       \font\mbfs = cmmib10 \font\mbfss = cmmib10 scaled 833
\font\msybf = cmbsy10 scaled\magstep1
       \font\msybfs = cmbsy10 \font\msybfss = cmbsy10 scaled 833
\textfont1 = \twelvei
       \scriptfont1 = \twelvei \scriptscriptfont1 = \teni
       \def\mit{\fam1 }
\textfont9 = \mbf
       \scriptfont9 = \mbfs \scriptscriptfont9 = \mbfss
       \def\bmit{\fam9 }
\textfont10 = \msybf
       \scriptfont10 = \msybfs \scriptscriptfont10 = \msybfss
       \def\bmsy{\fam10 }

\def\etal{{\it et al.~}}
\def\eg{{\it e.g.,~}}
\def\ie{{\it i.e.,~}}
\def\lsim{\raise0.3ex\hbox{$<$}\kern-0.75em{\lower0.65ex\hbox{$\sim$}}}
\def\gsim{\raise0.3ex\hbox{$>$}\kern-0.75em{\lower0.65ex\hbox{$\sim$}}}

\title{Cluster Mergers and Non-Thermal Phenomena:
A Statistical Magneto-Turbulent Model}

\author{R. Cassano$^{1,2}$ \& G. Brunetti $^{2}$}
\address{$^{1}$ Dipartimento di Astronomia ,Univ. di Bologna, Italy}
\address{$^{2}$ Istituto di Radioastronomia del CNR ,Bologna, Italy}

%



\abstract{
With the aim to investigate the statistical properties and the connection 
between thermal and non-thermal properties of the ICM in galaxy clusters, 
we have developed a statistical magneto-turbulent model 
which describes, at the same time, the evolution of the thermal and 
non-thermal emission from galaxy clusters.
In particular, starting from the cosmological evolution of clusters, we follow 
cluster mergers, calculate the spectrum of the magnetosonic waves generated in the 
ICM during these mergers, the evolution of relativistic electrons and the resulting 
synchrotron and Inverse Compton spectra. We show that the broad band (radio and 
hard x-ray) non-thermal spectral properties of galaxy clusters can be well 
accounted for by our model for viable values of the parameters (here we adopt a EdS cosmology).}


\maketitle

\section {Introduction}
The most important evidences for non-thermal phenomena 
in galaxy clusters comes from the synchrotron radio emission diffused on Mpc 
scales (Radio Halos (RH)) observed in a growing number of massive clusters (\eg Feretti 2003) and, 
more recently, from the hard X-ray tails (HXR tails) in excess of the thermal
bremmstrahlung spectrum detected in a few cases (Fusco-Femiano \etal 2004). 
RHs are usually found in merging clusters (e.g, Schucker \etal 2001) 
and the detection rate of RHs shows a abrupt increase with increasing 
the X-ray luminosity and mass of the host clusters: about 30-35\% of the galaxy clusters with  
X-ray luminosity larger than $10^{45}$ erg $s^{-1}$ show diffuse non-thermal 
radio emission (Giovannini \& Feretti 2002). This observations suggest a strong 
connection between the existence of non-thermal components and the dynamical
properties of the thermal ICM.\\ 
A promising possibility to expalain giant RHs is given by
the presence of relativistic electrons reaccelerated by MHD turbulence developed 
during cluster mergers in the ICM (Brunetti 2003; Brunetti 2004 these proceedings).
Up to now there are no theoretical works which investigate if the turbulent
reacceleration is able to reproduce the statistical properties of
RHs in the framework of a hierarchical scenario of cluster
formation. Thus, with the aim to investigate the statistical properties and 
the connection between thermal and non-thermal phenomena in galaxy clusters, 
we have developed a statistical magneto-turbulent model. 
In our work, we model the formation of RHs and HXR tails in a 
self-consistent approach which follows, at the same time, the evolution of 
the thermal properties of the ICM and the triggering and evolution of 
the non--thermal phenomena in the framework of the magneto-turbulent 
re-acceleration class of models. In particular, we follow the formation 
and evolution of clusters of galaxies, the generation of merger-driven 
turbulence in the cluster volume and the acceleration and time--evolution 
of the relativistic particles, and the related non-thermal emission.
The goals of our model are:
\begin{itemize}
\item[1] to check if cluster turbulence generated during 
mergers may be able to drive efficient particle acceleration processes in the ICM.
\item[2] to investigate, in the framework of the turbulent-acceleration 
hypotesis, if the hierarchical formation process of galaxy clusters can naturally
account for the observed statistics of RHs.
\end{itemize}
In the following we describe our model and investigate the point (1); 
a more detailed discussion is presented in Cassano \& Brunetti 2004. 
The second point and the Luminosity Function of RHs 
expected from this model are discussed in Cassano, Brunetti, Setti, these proceedings.  

\section {The magneto-turbulent model}

\subsection {Cluster formation}

The evolution and formation of galaxy clusters is computed making use of     
a relatively simple semi-analytical procedure based on hierarchical  
Press \& Schechter 1974 (PS) theory of cluster formation. 
We use the extended PS formalism developed by  Lacey \& Cole (1993)  
which gives the probability that a parent cluster of mass $M_{1}$ at time $t_{1}$ 
had a progenitor of mass $M_2 \rightarrow M_2 + dM_2$  at some earlier time $t_2$ , 
with $M_1 > M_2$ and $t_1> t_2$ (e.g., Lacey \& Cole 1993, Randall, Sarazin \&
Ricker 2002):

\begin{eqnarray}
\lefteqn{P(M_2, t_2|M_1, t_1)dM_2 =
\frac{1}{\sqrt{2\pi}}
\frac{M_1}{M_2}
\times } \\
& & \frac{\delta_{c2}-\delta_{c1}}{(\sigma_{2}^2-\sigma_{1}^2)^{3/2}}
\left| \frac{d\sigma_{2}^2}{dM_2} \right|
\exp\left[-\frac{(\delta_{c2}-\delta_{c1})^2}{2(\sigma_{2}^2-\sigma_{1}^2)}
\right] dM_2 ,
\nonumber
\end{eqnarray}

Where $\sigma(M)$  is the rms density fluctuation within a sphere of mean mass M and 
$\delta_c(z)$ is the critical linear overdensity for a region to collapse at redshift z.\\ 
Adopting a suitable change of variable (M with $S\equiv\sigma^{2}(M)$ and t with 
$x\equiv\delta_c(t)$) Eq.(1) is replaced by:

\begin{equation}
K(\Delta S, \Delta x) d \Delta S = \frac{1}{\sqrt{2 \pi}}
        \frac{\Delta x}{(\Delta S)^{3/2}}
        \exp \left[- \frac{(\Delta x)^2}{2 \Delta S} \right] d \Delta S
\, .
\label{mergermergerK}
\end{equation}

Thus the probability that a merger with a given $\Delta S$ (i.e. $\Delta M$) occurs at 
a given time can be calculated by making use of the cumulative distribution of 
subclusters masses:

\begin{eqnarray}
{\cal P}( <\Delta S, \Delta x ) =
\int_{0}^{\Delta S} K (\Delta S^{\prime}, \Delta x)
\, d\Delta S^{\prime}
=
\nonumber\\
{\rm erfc} \left( \frac{\Delta x}{\sqrt{2 \Delta S}} \right) \, ,
\label{mergermergerC}
\end{eqnarray}

where erfc() is the complementary error function. 

\begin {figure}[t]
\vskip 0cm
\centerline{\epsfysize=10cm\epsfbox{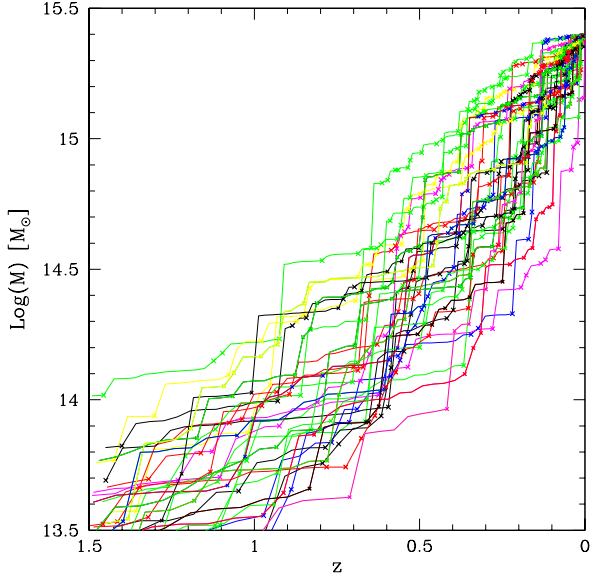}}
\vskip -2.3cm
\label{fig1}
\caption{Example of Merger Trees obtained from our Monte Carlo simulations in a EdS cosmology
for a cluster with a present day mass $M_o=2.5\times 10^{15}M_{\odot}$; each 
cross marks a merger event with $M_{2}\geq 10^{13} M_{\odot}$}
\vskip -0.2cm
\end{figure}

We employ a Monte Carlo procedure which selects a random number, r, 
in the range 0-1, and determines the corresponding value of $\Delta S$ solving
numerically the equation: ${\cal P}( < \Delta S, \Delta x ) = r$. 
The value of $S_2$ of the progenitor is  given by $S_2=S_2+\Delta S$ and the 
mass by $\sigma^2 ( M_2 ) = S_2$. The mass of the other subcluster is 
$\Delta M = M_1-M_2$. This procedure is thus iterated until the mass 
of the main cluster drops below a critical value or when a maximum redshift 
of interest is reached. An example of merger tree obtained from our
 procedure is illustrated in Fig.1.

\begin {figure}[t]
\vskip 0cm
\centerline{\epsfysize=10cm\epsfbox{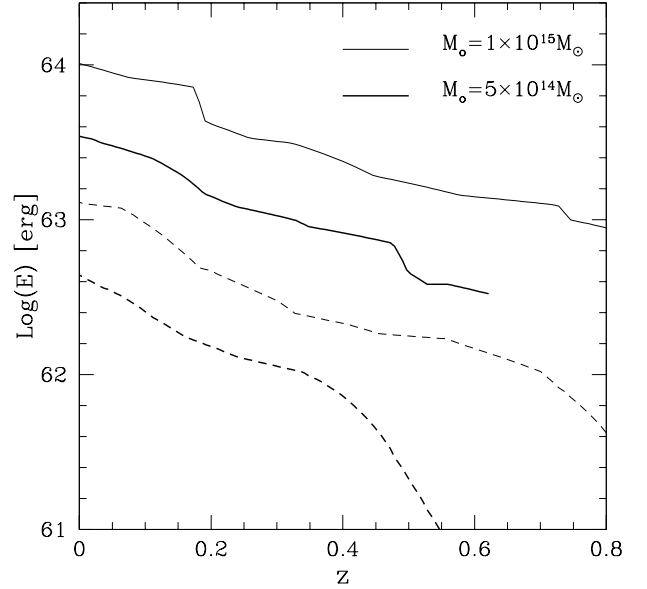}}
\vskip -2.5cm
\label{fig2}
\caption{Evolution of the thermal energy (solid lines) injected in fluid turbulence 
(dashed lines) in typical galaxy clusters. The thin lines are for a cluster with present 
time mass $M_o=10^{15}M_{\odot}$ and the tick lines are for a cluster with present time 
mass $M_o=5\times10^{14}M_{\odot}$.} 
\vskip -0.2cm
\end{figure}

\subsection {Turbulence in galaxy clusters}

The turbulence in the ICM is supposed to be injected during cluster mergers. 
The energetics of the turbulence injected in the ICM is calibrated with the 
$PdV$ work done by the infalling subclusters in passing through the volume 
of the most massive one. This energetics is given by
$E_{t}\simeq<\rho>_{ICM}v_{i}^{2}V_{t}$ where $V_{t}$ is the volume sweaped by
the infalling subhalos which is estimated following standard recipes based
on {\it Ram Pressure Stripping} (e.g., Fujita,Takizawa, Sarazin 2003; 
Cassano \& Brunetti 2004) and $v_{i}$ is the impact velocity between 
the two colliding subclusters. Turbulence driven by a given merger 
is assumed to be injected and then dissipated within a crossing time, 
$\tau_{cros}$. We assume that a fraction $\eta_{t}$ of 
the energy of the turbulence developed during these mergers is channeled into 
{\it fast magnetosonic waves} (MS waves), $E_{MS}=\eta_{t}E_{t}$.

\begin {figure*}[t]
\vskip 0cm
\centerline{\epsfysize=12.8cm\epsfbox{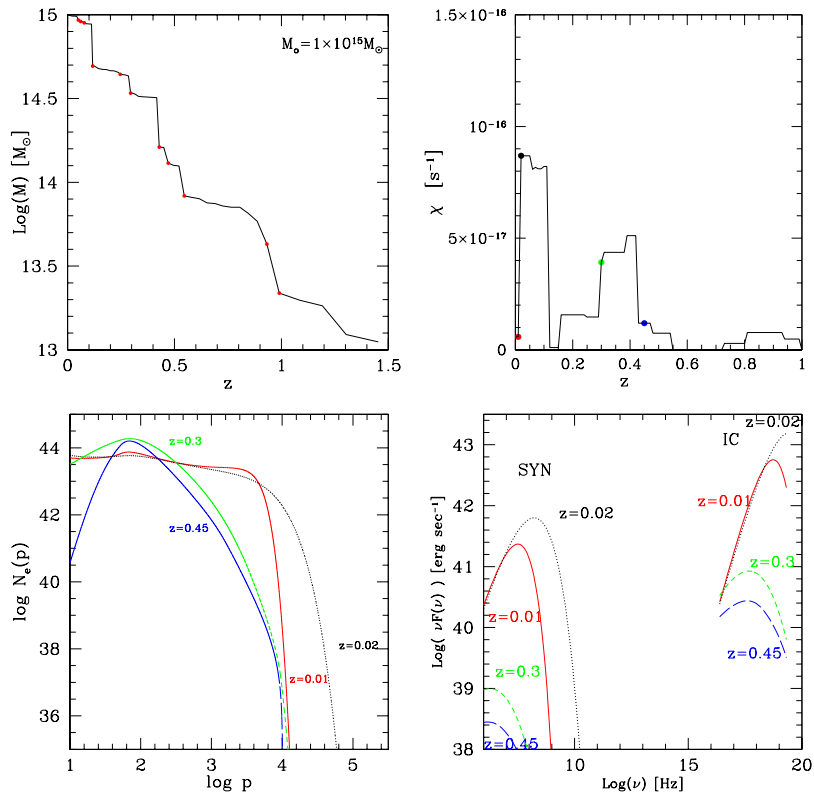}}
\vskip -2.0cm
\label{fig3}
\caption{ a) a singol merger tree for a cluster with present-day mass of
$M_{o}=1\times 10^{15}M_{\odot}$, each dot represents a merger event with
$M_{2}\geq 10^{13} M_{\odot}$; b) the time-evolution (with redshift) of the electron acceleration coefficient ($\tau_{acc}=\chi_{acc}^{-1}$) 
during the cluster formation; c) the evolution of the electron spectrum at different relevant 
times (see panel) in the same cluster; c) the corresponding radio (synchrotron) 
and hard-X ray (IC) emission. Calculations are performed assuming:
$R_{H}=500 h_{50}^{-1} Kpc$, $\eta_{t}=0.26$, $\eta_{e}=0.003$ and $<B>=0.5
\mu$G.}   
\vskip -0.2cm
\end{figure*}

We use these waves since their damping rate and time evolution basically depend on the 
properties of the thermal plasma which are provided by our merger trees for
each synthetic cluster. As a relevant example, in Fig.2 we report the cosmological evolution of the thermal 
energy of galaxy clusters with different masses, together with the total energy injected 
in form of turbulence in the ICM. The energy in form of turbulence is calculated by 
integrating the contributions from all the merger events experienced during
the cluster life. The thermal energy of the 
considered clusters, calculated assuming the observed {\it M-T} relation 
(e.g., Nevalainen et al. 2000), increases from about
$10^{62}$erg at $z \sim 1$ to a few $10^{64}$erg at the present
epoch depending on the mass of the cluster. As it should be, we note that
the energy budget injected in turbulence during cluster formation is well below the
thermal energy ($\sim15\%$ of it), which is also in agreement with 
recent numerical simulations (Sunyaev et al. 2003) and with very recent observational claims 
(Schuecker et al. 2004). 
Finally, as reasonably expected, the energy injected in turbulence
calculated with our approach is found to roughly
scale with the thermal energy of the clusters.\\
Under the physical conditions in the ICM, the spectrum of MS waves 
injected during each merger 
can be estimated as (Cassano \& Brunetti 2004):     
\begin{eqnarray}
W_{\rm k}
\simeq
\frac{I(\rm k)}
{\Gamma_{\rm th,e}(k)}
\label{ms-stationary}
\end{eqnarray}

where $\Gamma_{th,e}$ is the damping coefficient with thermal electrons and $I_k$ 
is the spectrum of the injection rate of MS waves (we assume a Kolmogorov
spectrum, $I(k)\propto k^{-5/3}$ ). We are interested in the energy of 
the turbulence injected in a volume typical of a RH ($V_{H}=4\pi
R_H^3/3$). The normalization of $I(k)$ is thus given by:          

\begin{equation}
V_H\int_{k_{0}} I(k)dk \simeq\frac{\eta_t E_{t}}{\tau_{cros}},
\label{Ioint}
\end{equation}

where $k_{0}$ is the wavenumber correspondent to the maximum scale which is
fixed at the stripping radius for each merger event (Cassano \& Brunetti 2004). 

\subsection {Particle Acceleration}

We assume the presence of relativistic electrons in the ICM which are 
continuously injected by AGNs, Galactic Winds, and/or merger shocks. 
Given the calculated (Sect. b)) spectrum of MS waves and the physical conditions 
in the ICM (Sect. a)), we compute the time evolution of relativistic electrons at 
each time step by solving a Fokker-Planck equation including the  
electron acceleration due to MS waves:

\begin{eqnarray}
{{\partial N(p,t)}\over{\partial t}}=
{{\partial }\over{\partial p}}
\left[
N(p,t)\left(
{ \big| {{dp}\over{dt}} \big|}_{\rm rad} +
{\big| {{dp}\over{dt}} \big| }_{\rm c}
-{2\over{p}} D_{\rm pp}
\right)\right] +
 &  &         \nonumber \\
{{\partial }\over{\partial p}}
\left[
D_{\rm pp}
{{\partial N(p,t)}\over{\partial p}}
\right] +
Q_e(p,t)
\label{elettroni}
\end{eqnarray}

$(dp/dt)_c$ is the Coulomb losses term due to the interation with the thermal plasma 
(proportional to the density of the ICM) and $(dp/dt)_{rad}$ is the term of radiative 
losses due to the synchrotron and IC scattering off the CMB photon.
For the injection rate of relativistic electrons we adopt a power law spectrum
up to a maximum momentum:

\begin{equation}
Q_e(p)=K_e p^{-s}
\label{Qe}
\end{equation}

We normalize the injection rate by assuming that the total energy
injected in relativistic electrons during the cluster life is a fraction,
 $\eta_e$, of the total thermal energy of the cluster at z=0.

The electron diffusion coefficient in 
the momentum space due to the interaction with the MS waves (\eg Eilek 1979;
Cassano \& Brunetti 2004) is given by:

\begin{equation}
D_{\rm pp}(p,t)\simeq
4.45 \, \pi^{2}\,{{ v_M^2 }\over{c}}
{{p^2}\over{B^2}}
\int_{k_{min}}^{k_{max}}
k {\cal W}^{B}_k(t)
dk
\label{dppms}
\end{equation}

The acceleration time scale is $\tau_{acc}^{-1}=\chi\simeq 2 D_{pp}/p^2$  and thus 
the systematic energy gain of electrons due to MS waves is $(dp/dt)_{acc}=\chi p$. 
The coefficient of electron acceleration at redshift z is obtained by combining the effect due 
to the MS waves injected during the mergers which contribute to the
turbulent spectrum at z, one has (Cassano \& Brunetti 2004):

\begin{eqnarray}
\chi(z)
\simeq {{ 2.23 \times 10^{-16} \eta_t}
\over{ (R_H/500{\rm kpc})^3}}
\sum_{j}
\Bigg[
\Big(
{ {M_{1} +M_{2} }\over{
2\times 10^{15} M_{\odot}}}
\,\,
{{ 2.6 {\rm Mpc} }\over
{ R_{1}}}
\Big)^{3/2}
\nonumber\\
 \times
{{(r_s/500 {\rm kpc})^2}\over{(kT/7 {\rm keV})^{1/2}}}
\Bigg]_{j}
\label{dppz}
\end{eqnarray}

where $R_H$ is the radius of the RH and $r_s$ is the stripping 
radius of the subclumps.\\

In Fig.(3) we report an example of our results for a cluster with present-day mass 
$M_{o}=10^{15}M_{\odot}$. In Fig.(3a) we report the merger history (merger
tree) which is obtained by using extended P\&S formalism. Given the merger tree, we 
calculate the turbulence injected during cluster mergers and assume that 
a fraction $\eta_{t}$ is channeled in MS waves ($\eta_{t}\simeq 0.26$ is
adopted in this case). The coefficient of electron-acceleration is calculated 
considering all the merger events which contribute to the injection of turbulence at redshift 
z in the considered cluster (Fig.(3b)). The evolution with redshift of the electron 
spectra (Fig.~(3c)) is obtained solving the Fokker-Planck equation (Eq.(\ref{elettroni})) 
which accounts for the relevant energy losses and the acceleration of the relativistic 
electrons due to MS waves. In Fig.(3) we assume that the energy injected in 
relativistic electrons during the cluster life is a fraction ($\eta_{e}=0.003$) of 
the present energy of the thermal pool and calculate the synchrotron and IC emission 
spectra at different redshift (Fig.(3d)) from a region of $R_{H}\sim 500$ kpc $h_{50}^{-1}$ 
and adopting an average magnetic field of $<B> \simeq 0.5$ $\mu$G.\\
The main goal of Fig.(3) is to show that the typical observed radio, $L_{R}\simeq10^{40}-10^{41}$ erg 
$s^{-1}$ (\eg, Feretti 2003) and hard X-ray, $L_{HX}\simeq10^{43}-10^{44}$ erg $s^{-1}$ 
(\eg, Fusco-Femiano \etal 2003) luminosities can be obtained in $ M\gsim
10^{15} M_{\odot}$ clusters during major merger events, provided that 
a fraction of the cluster thermal energy (of the order of $3-5\%$) is channeled into MS waves.
The additional requirement is that a population of relativistic electrons to
be reaccelerated is present in the ICM. We find that a total energy injected
into these electrons of a few $10^{-4}-10^{-3}$ times the present energy of the
thermal pool is sufficient.

\acknowledgements{G.B. and R.C. acknowledge partial support from CNR grant
CNRG00CF0A.}


\begin{references}
\reference{} Brunetti G., 2003, in 'Matter and Energy in Clusters
of Galaxis', ASP Conf. Series, vol.301, p.349, eds. S. Bowyer and C.-Y. Hwang.
\reference{} Cassano R., Brunetti G., 2004 submitted to MNRAS.
\reference{} Eilek J.A., 1979, ApJ 230, 373
\reference{} Feretti L., 2003, in 'Matter and Energy in Clusters
of Galaxis', ASP Conf. Series, vol.301, p.143,
eds. S. Bowyer and C.-Y. Hwang.
\reference{} Fujita Y., Takizawa M., Sarazin C.L., 2003,
ApJ 584, 190.
\reference{} Fusco-Femiano R., Orlandini M., Brunetti G.,
Feretti L., Giovannini G., Grandi P., Setti G., 2004, ApJ 602, 73.
\reference{} Giovannini G., Tordi M., Feretti L.,
1999, NewA 4, 141.
\reference{} Lacey, C., Cole S., 1993, MNRAS 262, 627.
\reference{} Nevalainen, J., Markevitch, M., Forman, W., 2000, ApJ 532, 694.
\reference{} Press W.H., Schechter P., 1974, ApJ 187, 425.
\reference{} Randall S.W., Sarazin C.L., Ricker P.M., 2002, ApJ 577, 579. 
\reference{} Schuecker P., B$\ddot{o}$hringer H., Reiprich, T.H., Feretti L.,
2001, A\&A 378, 408.
\reference{} Schuecker P., Finoguenov A., Miniati F., Boehringer H.,
Briel U.G., 2004, A\&A submitted; astro-ph/0404132
\reference{} Sunyaev, R. A., Norman, M. L., Bryan, G. L.,  2003,
Astronomy Letters, vol. 29, p. 783-790.



\end{references}
\end{document}